\documentclass[reprint,amsmath,amssymb,aps,pra,superscriptaddress,twocolumn]{revtex4-2}

\usepackage{graphicx}
\usepackage{siunitx}
\usepackage[version=4]{mhchem}

\begin{document}


\title{Feasibility analysis of a proposed test of quantum gravity via novel optical magnetometry in xenon}


\author{J.~Maldaner}
\email{maldaner@ualberta.ca}
\affiliation{Department of Physics, University of Alberta, Edmonton, Alberta T6G 2E1, Canada}
\author{M.~Fridman}
\affiliation{Department of Physics and Astronomy, University of Lethbridge, Lethbridge, Alberta T1K 3M4, Canada}
\author{S.~Das}
\affiliation{Department of Physics and Astronomy, University of Lethbridge, Lethbridge, Alberta T1K 3M4, Canada}
\author{G.~Porat}
\email{gporat@ualberta.ca}
\affiliation{Department of Physics, University of Alberta, Edmonton, Alberta T6G 2E1, Canada}
\affiliation{Department of Electrical and Computer Engineering, University of Alberta, Edmonton, Alberta T6G 1H9, Canada}


\date{\today}

\begin{abstract}
We present an analysis of the sensitivity limits of a proposed experimental
search for quantum gravity, using a novel approach based on optical
magnetometry in the noble gas isotope $\text{\ensuremath{^{129}}Xe}$.
The analysis relies on a general uncertainty principle model that
is consistent with most formulations of quantum gravity theory, where
the canonical uncertainty relations are modified by a leading-order
correction term that is linear in momentum. In turn, this correction
modifies the magnetic moment of the spin-polarized $\ce{^{129}Xe}$
atoms that are immersed in a magnetic field in the proposed experiment,
which results in a velocity-dependent variation of their Larmour frequency,
that is detected via two-photon laser spectroscopy. The thermal distribution
of atomic velocities, in conjunction with the Doppler effect, is used
to scan the interrogating laser over different atomic velocities,
and search for a corresponding variation in their Larmor frequencies.
We show that the existing bounds on the leading-order quantum gravity
correction can be improved by $10^{7}$ with existing technology,
where another factor of $10^{2}$ is possible with near-future technical
capabilities.
\end{abstract}


\maketitle
\section{Introduction}

Quantum gravity (QG), the union of general relativity and quantum
mechanics, is one of the most fundamentally significant outstanding
problems in theoretical physics. The study of QG has resulted in a
number of distinct proposed formulations including superstring theory,
loop quantum gravity, and doubly special relativity, to name a few
\cite{Isham-1993,Ali-2011}. The generalized uncertainty principle
(GUP) is consistent with all the above forms of QG theory and predicts modifications
in several areas of quantum mechanics, allowing for experimental searches
for QG effects without the need to conform to a specific formulation
\cite{Bosso-2017,Jun-Li-2021}. While most GUP formulations consider
only a correction that is quadratic in momentum, recently a more general
model has been developed, which includes both linear and quadratic
corrections \cite{Bosso-2017,Ali_Das_Vagenas-2009}. For example,
in one dimension, the GUP is given by 
\begin{equation}
\text{\ensuremath{\Delta x\Delta p\ge\frac{\hbar}{2}\left[1-2\frac{\alpha_{0}}{M_{p}c}\left\langle p\right\rangle +4\frac{\beta_{0}}{M_{p}^{2}c^{2}}\left\langle p^{2}\right\rangle \right],}}
\end{equation}
where $x$ is position, $p$ is momentum, $\alpha_{0}$ and $\beta_{0}$
are the linear and quadratic dimensionless GUP parameters, respectively,
$c$ is the speed of light, and $M_{p}=\sqrt{\hbar c/G}$ is the Planck
mass, where $\hbar$ is the reduced Planck's constant, $G$ is the
gravitational constant, and $M_{p}c$ is the Planck momentum.

The GUP parameter $\alpha_{0}$ defines a minimum possible length
$\ell_{min}\propto\alpha_{0}\ell_{P}$, where $\ell_{P}=\sqrt{\hbar G/c^{3}}$
is the Planck length. It is generally agreed that $\ell_{min}$ must
be between the Planck length $\ell_{P}\sim10^{-35}\SI{}{m}$ and the
electroweak length $\ell_{EW}\sim10^{-18}\SI{}{m}$ \cite{Das-2008,Majumder-2011},
where the latter is the scale in which the Large Hadron Collider operates
and in which no minimum length has been observed \cite{Ali-2012}. This places
the commonly agreed upon upper bound of the linear GUP parameter at
$\alpha_{0}\leq10^{17}$. As for the quadratic correction term, different
methods have placed experimental upper bounds on $\beta_{0}$, ranging
from $10^{21}$ to $10^{6}$ \cite{Das-2008,Ghosh-2013,Das-2022,Bushev-2019}.

This paper is published in conjunction with the paper by Fridman et
al. \cite{Fridman-2023}, which analyzes the effects of the linear and
quadratic GUP corrections on the nuclear magnetic moment of an atom,
and outlines an experiment for detecting these effects through Larmor
frequency measurements in a novel optical magnetometer utilizing direct
(ultraviolet) optical access to atomic transitions in the $\text{\ensuremath{^{129}}Xe}$
noble gas atom. An optical magnetometer measures the spin precession
frequency of an ensemble of spin-polarized atoms, i.e., the Larmor
frequency. Without QG correction, this frequency is given by $\omega_{B,0}=\gamma B$,
where $\gamma$ is the gyromagnetic ratio of the atom and $B$ is
the magnetic field strength. When the GUP QG correction is introduced
and we include corrections up to the $\mathcal{O}(m^{2}v^{2})$ term,
the Larmor frequency becomes $\omega_{B}(v)=\omega_{B,0}\left(1-\alpha_{0}mv/M_{p}c+\beta_{0}m^{2}v^{2}/M_{p}^{2}c^{2}\right)$,
where $m$ is the atom's mass and $v$ is the atom's velocity \cite{Ali-2011}.
Therefore, if the Larmor frequency in an optical magentometer is experimentally
shown to depend on the velocity of the magnetometer's atoms, it will
be an indication of QG effects. Conversely, if no dependence is shown,
the measurement's sensitivity can set upper bounds on $\alpha_{0}$
and $\beta_{0}$.

Here, we carry out an analysis of the sensitivity limits of the proposed
experiment. We consider a thermal atomic ensemble and include only
the first-order correction term parameterized by $\alpha_{0}$ and
neglect $\beta_{0}$, for the following reasons. First, even if an
ensemble of oganesson atoms, the heaviest discovered element, is at
room temperature, the ratio of the most probable momentum (in a Boltzmann-Mawell
distribution \cite{Demtroder2002}) of such atoms to the Planck momentum
is $mv/M_{p}c=\sqrt{2m_{\ce{^{294}Og }}k_{B}T}/M_{p}c\sim10^{-23}$,
where $m_{^{294}\ce{Og}}$is the mass of oganesson, $k_{B}$ is Boltzman's
constant, and $T=\SI{300}{K}$ is room temperature. Therefore, as
long as $\beta_{0}/\alpha_{0}\ll10^{23}$ the quadratic component
in Eq. \ref{eq:qg_correction} can be neglected. Given that the existing
bounds on $\beta_{0}$ range from $10^{21}$ to $10^{6}$, the quadratic
component is negligible at least for $\alpha{}_{0}>10^{-2}$, and
possibly even for $\alpha_{0}>10^{-17}$. Hence, there is a wide range
of possible values of interest of $\alpha_{0}$, i.e. $10^{-2}<\alpha_{0}<10^{17}$,
where the quadratic term is negligible. Therefore, henceforth we denote
the QG correction
\begin{equation}
\mathcal{C}(v)=\alpha_{0}\frac{mv}{M_{p}c},\label{eq:qg_correction}
\end{equation}
so that the QG-corrected Larmor frequency is
\begin{equation}
\omega_{B}\left(v\right)=\omega_{B,0}\left(1-\mathcal{C}\left(v\right)\right).\label{eq:QG_Larmor}
\end{equation}

In our analysis, we consider existing or near-future technological
capabilities, where we propose measuring the QG-induced Larmor frequency
variation across the velocity distribution of a thermal ensemble of
hyperpolarized $^{129}\text{Xe}$ atoms via two-photon optical magnetometry,
with some similarity to the magnetometer proposed by Altiere et al.
\cite{Altiere_PRA_Xe_two-photon}. In section \ref{sec:Proposed-Experimental-Methodolog}
we detail the experimental methodology and derive the signal and noise
models of the optical magnetometry experiment. In section \ref{sec:Parameter-Values}
we outline the state of the art in magnetic field uniformity, deep
ultraviolet (DUV) laser power and frequency stability, and techniques
to maximize the spin polarization decay time of $\ce{^{129}Xe}$.
We use these data to constrain experimental parameters in our model.
Finally, in section \ref{sec:results}, we optimize the values of
the remaining unconstrained experimental parameters, calculate the
corresponding sensitivity and bound on $\alpha_{0}$, and discuss
the results, where we also consider and reasonable extrapolations
of technical capabilities. Our analysis shows that it is feasible
for this method's sensitivity to reveal QG signatures corresponding
to $\alpha_{0}\gtrsim10^{9}$ with existing technology, where key
factors are a high-power, tunable CW or high pulse repetition rate
frequency comb laser in the DUV, and a high degree of spatial uniformity
of the magnetic field.

\section{Proposed Experimental Methodology and Sensitivity Estimation}

\label{sec:Proposed-Experimental-Methodolog}

\subsection{Experimental Methodology\label{subsec:Experimental-Methodology}}

\begin{figure}
\includegraphics[width=0.45\textwidth]{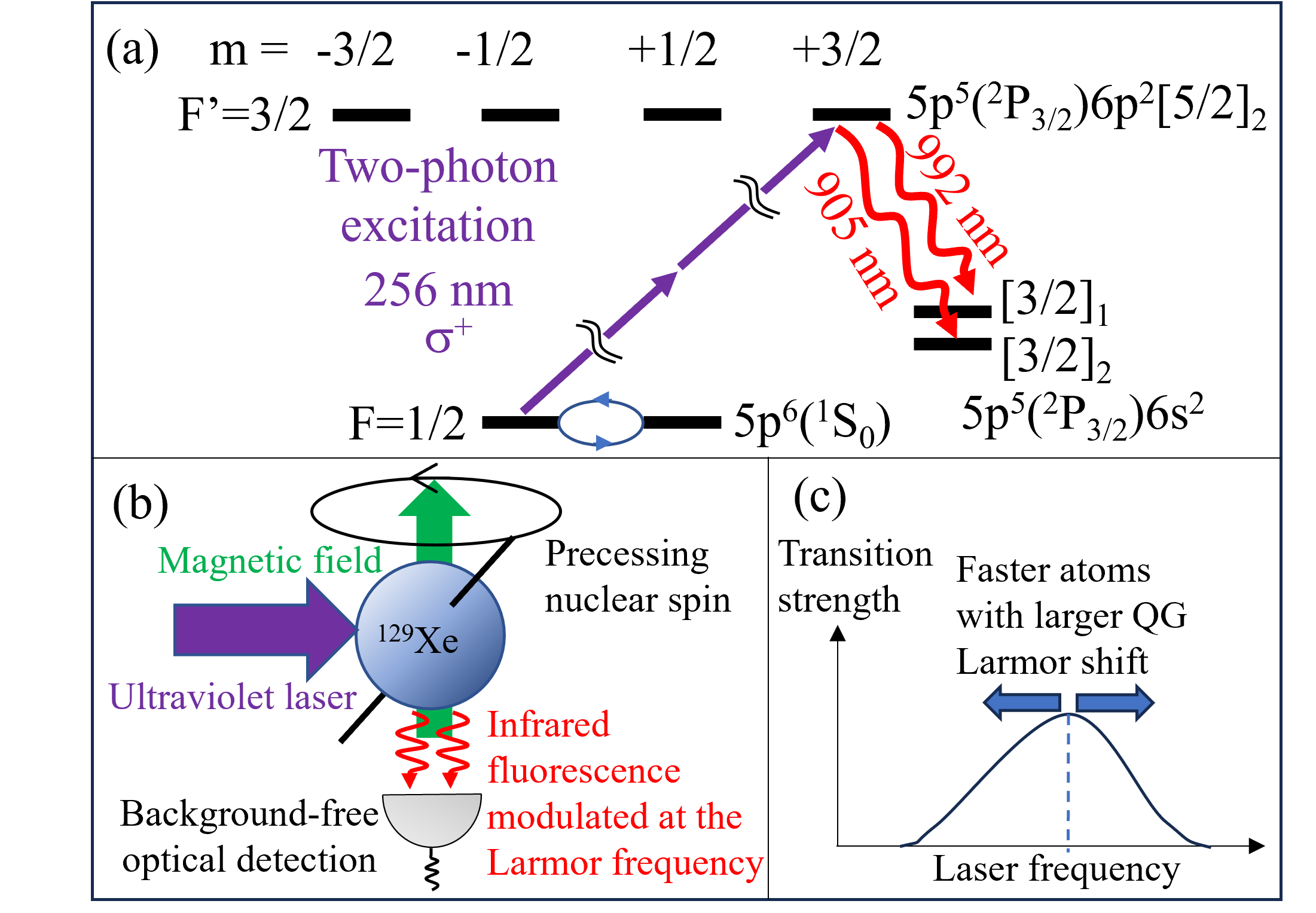}
\caption{\label{fig:experiment_and_energy_diagram}(a) The
partial energy level diagram of $\ce{^{129}Xe}$. A circularly-polarized
ultraviolet laser drives spin-selective two-photon excitation in hyperpolarized
$\ce{^{129}Xe}$ atoms, where the excited state emits infrared light
via radiative decay. The nuclear spin of the atoms precesses around
an applied magnetic field, which is perpendicular to the laser propagation
direction. Correspondingly, the ground state population coherently
oscillates between the two magnetic sublevels, resulting in oscillation
of the emitted infrared fluorescence. The latter is detected for extraction
of the precession (Larmor) frequency.  (b) Illustration of the proposed experiment. (c) Tuning the laser to different
frequencies in the Doppler-broadened profile of the two-photon transition
corresponds to different atom velocities, which in turn are associated
with different QG shifts of the Lamor frequency.}
\end{figure}

The principle of our proposed experiment is illustrated in Fig. \ref{fig:experiment_and_energy_diagram}
and follows the general principles of optical magnetometers \cite{Budker-2007}.
We assume a hyperpolarized atomic gas of $\ce{^{129}Xe}$ atoms, i.e.,
where the atoms are in the same magnetic sublevel of the ground state.
These atoms are placed in a magnetic field that is perpendicular to
their spin orientation, i.e., perpendicular to the quantization axis
used in the energy level diagram depicted in Fig. \ref{fig:experiment_and_energy_diagram}
(a). Therefore, the collective atomic spin precession about the magnetic
field is manifested as a coherent oscillation of the atomic state
between the ground state's two magnetic sublevels, at the Larmor frequency.
An ultraviolet laser beam, with $\sigma^{+}$ circular polarization
and wavelength of $\SI{256}{nm}$, propagates through the gas in a
direction perpendicular to the magnetic field. This laser light drives
a two-photon transition from the ground state $m=-1/2$ sublevel to
the $\ce{5p^{5}(^{2}P_{3/2})6p^{2}[5/2]_{2}},\,m=+3/2$ excited state.
Note that, under these conditions, there is no dipole-allowed two-photon
transition from the the ground state $m=+1/2$ sublevel. Therefore,
since the atomic population of the ground state $m=-1/2$ sublevel
oscillates at the Larmor frequency, so does the excited population.
The excited atoms then decay to either the $\ce{5p^{5}(^{2}P_{3/2})6s^{2}[3/2]_{1}}$or
$\ce{5p^{5}(^{2}P_{3/2})6s^{2}[3/2]_{2}}$state, by emitting near-infrared
(NIR) radiation, before decaying back to the ground state. Hence,
the amplitude of the NIR radiation follows the excited population
and oscillates at the Larmor frequency. In our experiment, the NIR
radiation is detected and the resulting signal is analyzed for extracting
the Larmor frequency, which carries information on the QG correction
as expressed by Eqs. \ref{eq:qg_correction} and \ref{eq:QG_Larmor}.

The variation in the Larmor frequency, $\omega_{B}$, depends on the
velocity of the xenon atoms, with greater velocity producing a greater
shift. Therefore, identical atoms must be traveling at different velocities
through the same magnetic field to produce different shifts of their
Larmor frequency, for these shifts to be detected. Atoms in a thermal
ensemble move with a wide range of velocities, described by the Maxwell-Boltzmann
distribution. Due to the atoms' motion, their transition resonance
frequencies are Doppler shifted, resulting in a Gaussian distribution
of resonance frequencies that is broader than the natural linewidth
\cite{Demtroder2002}. When sweeping the laser frequency across the
Doppler broadened profile, where the laser linewidth is much smaller
than the Doppler width, xenon atoms with the corresponding velocity
are excited (see Fig. \ref{fig:experiment_and_energy_diagram} (c)).
The relationship between the laser angular frequency $\omega_{l}$
in the lab frame of reference, the resonant angular frequency $\omega_{\ce{Xe}}$
in the xenon atom's frame of reference, and the atom's velocity $v$
is \cite{Hecht-2014}
\begin{equation}
v\left(\omega_{l}\right)=c\left(1-\frac{\omega_{\text{Xe}}}{\omega_{l}}\right).\label{eq:lambda_and_velocity}
\end{equation}
As the laser probes atoms with different velocities, this corresponds
to different values of $\mathcal{C}\left(\omega_{l}\right)=\mathcal{C}\left(v\left(\omega_{l}\right)\right)$,
and therefore a different value of $\omega_{B}\left(\omega_{l}\right)=\omega_{B}\left(v\left(\omega_{l}\right)\right)$.
With adequate precision in the measurement of $\omega_{B}\left(\omega_{l}\right)$,
it is possible to detect the deviation caused by the QG effect. The
possibility of detecting the QG signature depends on the value of
$\alpha_{0}$, with larger values resulting in a larger QG effect.
The proposed experiment would then result either in a measurement
of $\alpha_{0}$ or provide an upper bound for its value. It should
be noted that while not all atoms that are probed at a specific $\omega_{l}$
have the same velocity. Rather, they have the same longitudinal velocity.
Due to the symmetry of the Maxwell-Boltzmann distribution, the contribution
of the transverse velocity components is the same for any excitation
frequency. Therefore, here they are ignored, since we are only aiming
to resolve the difference between Larmor frequencies at different
laser excitation frequencies.

We note that optical magnetometry is advantageous since the measurement
is performed \emph{in situ} and not at a distance as with other high-precision
magnetometers, e.g., superconducting quantum interference devices
(SQUIDs) \cite{Stolz-2022}. Also, SQUIDs need to be cryogenically
cooled, introducing significant technical challenges, while here we
propose a method with room temperature operation. The main advantages
of using hyperpolarized $\ce{^{129}Xe}$ are the two-photon electronic
transitions at frequencies that are low relative to other noble gases,
and very long coherence time of its nuclear spin (scale of hours)
at room temperature \cite{kilianFreePrecessionTransverse2007a,Anger-2008},
characteristic of noble gas isotopes with a nuclear spin of 1/2. The
latter stems from the fact that the only non-zero angular momentum
of the ground state of $\ce{^{129}Xe}$ is the nuclear spin of 1/2,
as the full electronic shell configuration has net zero electronic
spin and orbital angular momenta. In hyperpolarized $\ce{^{129}Xe}$,
the electrons have been pumped to a single magnetic sublevel of the
ground state, which means that the $\ce{^{129}Xe}$ atoms are nuclear
spin-polarized, and their nuclear spins can coherently precess under
the influence of an external magnetic field with very little perturbation
by the external environment. Another attractive feature of the proposed
scheme is the use of a UV-excited state that decays via NIR emission,
which provides a background-free signal that is easy to detect with
technologically mature low-noise detectors.

The figure of merit used in this paper is the fractional sensitivity
$S$, defined as $S=\delta\omega_{B}/\omega_{B}$, where $\delta\omega_{B}$
is the statistical uncertainty in the detected Larmor frequency $\omega_{B}$.
To detect the expected shift in the Larmor frequency due to QG, the
Larmor frequency must be measured at minimum at two laser frequencies
(preferably more), $\omega_{l,1}$ and $\omega_{l,2}$, with a fractional
sensitivity that satisfies $S\leq\left|\omega_{B}\left(\omega_{l,1}\right)-\omega_{B}\left(\omega_{l,2}\right)\right|/\omega_{B}\left(\omega_{\ce{Xe}}\right)=\left|\mathcal{C}\left(\omega_{l,1}\right)-\mathcal{C}\left(\omega_{l,2}\right)\right|$.
If one measurement is taken on resonance ($\omega_{l}=\omega_{\text{Xe}})$
and the other at a frequency corresponding to velocity $v$, then
to detect the QG signature we need $S\leq\mathcal{C}\left(\omega_{l}\right)$.
With this figure of merit we can determine the minimum detectable
value of $\alpha_{0}$, denoted $\alpha_{0,\text{min}}$. Then the
minimum possible detectable $\alpha_{0}$ is given by $\alpha_{0,min}=S\cdot\left(M_{p}c/mv\right)$
where both $S$ and $v$ depend on the laser frequency. We show in
Section \ref{sec:results} that, e.g., a fractional sensitivity on
the order of $10^{-14}$ is required for detecting $\alpha_{0}\approx10^{9}$.

The proposed experimental setup consists of a small, spherical cell
that contains xenon gas. The cell is contained within an apparatus
that generates an extremely uniform magnetic field, and is illuminated
by a DUV laser. The fluorescence emitted by the excited xenon atoms
is collected into an optical detector. More details and feasibility
estimation of required experimental conditions are discussed in Section
\ref{sec:Parameter-Values}. The essence of the experimental procedure
is as follows:
\begin{enumerate}
\item Prepare the xenon atoms in a spin-polarized state, e.g., via spin-exchange
optical pumping \cite{nikolaouNearunityNuclearPolarization2013a}.
\item Place the atoms in a magnetic field and illuminate them with circularly
polarized laser light that propagates perpendicularly to the magnetic field direction, and has a frequency near a resonance of a two-photon transition in xenon.
\item As the atoms precess in the magnetic field, get excited by the laser
and decay back to the ground state, measure their infrared emission,
which oscillates at the Larmor frequency.
\item Repeat steps 1-3 until the fractional sensitivity has reached the
desired level through statistical averaging.
\item Repeat steps 1-4 for different laser frequencies within the Doppler
broadened lineshape of the two-photon transition.
\end{enumerate}
In the following two sections we outline the fluorescence detection
model and derive an analytical expression for the fractional sensitivity
as a function of experimental parameters.

\subsection{Detected Signal Model}
Towards determining the fractional sensitivity, we develop an analytical
model for the infrared photon detection rate. We begin by determining
the rate at which the xenon atoms absorb the DUV laser light. The
two-photon absorption rate per atom is \cite{Saxon-1986}
\begin{equation}
W^{(2)}(\omega_{l})=\hat{\sigma}^{(2)}\left(\omega_{l}\right)\left(\frac{I}{\hbar\omega_{l}}\right)^{2},\label{eq:tw-photon-abs-rate-w-cross-section}
\end{equation}

where $\hat{\sigma}^{(2)}(\omega_{l})$ is the two photon cross-section,
$I$ is the intensity of the laser, $\omega_{l}$ is the angular frequency
of the laser radiation, and $\hbar$ is the reduced Planck's constant.

We express the Doppler-broadened two-photon cross-section with experimentally
determined parameters as (see Appendix \ref{sec:Two-photon-cross-section})
\begin{equation}
\begin{split}
\hat{\sigma}^{(2)}(\omega_{l}) = & \alpha_{\text{Xe}}(\hbar\omega_{l})^{2}\sqrt{\frac{\ln2}{\pi}}\frac{2}{\Delta\omega_{D}} \times \\
& \exp\left[-\ln2\left(\frac{2}{\Delta\omega_{D}}\right)^{2}\left(\omega_{l}-\omega_{\ce{Xe}}\right)^{2}\right],\label{eq:two-photon-cross-section}
\end{split}
\end{equation}
where $\alpha_{\ce{Xe}}$ is the two-photon absorption coefficient
and the full-width at half-maximum (FWHM) of the Doppler broadened
absorption profile is \cite{Demtroder2002} 
\begin{equation}
\Delta\omega_{D}=\frac{\omega_{\text{Xe}}}{c}\sqrt{\frac{8k_{B}T_{\text{Xe}}\ln2}{m_{\text{Xe}}}},\label{eq:doppler_FWHM}
\end{equation}
where $T_{\text{Xe}}$ is the temperature of the atomic gas, $k_{B}$
is the Boltzmann constant, $m_{\text{Xe}}$ is the mass of a $\ce{^{129}Xe}$
atom, $c$ is the speed of light, and $\omega_{\text{Xe}}$ is the
transition resonance angular frequency in the atom's frame of reference.
The two-photon transition rate per atom is therefore
\begin{equation}
\begin{split}
W^{\left(2\right)}\left(\omega_{l}\right) = & \alpha_{\text{Xe}}\sqrt{\frac{\ln2}{\pi}}\frac{2}{\Delta\omega_{D}}\left(\frac{2P_{l}}{\pi w_{0}^{2}}\right)^{2} \times \\ & \exp\left[-4\ln2\left(\frac{\omega_{l}-\omega_{\ce{Xe}}}{\Delta\omega_{D}}\right)^{2}\right],\label{eq:two-photon-abs-rate}
\end{split}
\end{equation}
where $P_{l}=I\pi w_{0}^{2}/2$ is the laser power, and $w_{0}$ is
the beam radius at $1/e^{2}$ of maximum intensity, where we assume
a Gaussian beam with peak intensity $I$.

By design of the experimental geometry, the beam size is approximately
constant throughout the interaction region, i.e., $b=2z_{R}\gg L$,
where $z_{R}=\pi w_{0}^{2}/\lambda$ is the Rayleigh range, $b$ is
the beam's confocal parameter, and $L$ is the interaction length.
Therefore, the interaction volume is $L\pi w_{0}^{2}$, and the ensemble
absorption rate is $W^{\left(2\right)}\left(L\pi w_{0}^{2}\right)n_{\text{Xe}}$,
where $n_{\text{Xe}}$ is the number density of $^{129}\text{Xe}$
atoms in the gas. We note that $n_{\text{Xe}}$ is related to the
temperature $T_{\text{Xe}}$ and pressure $p_{\text{Xe}}$ of the
gas through ideal gas law, $n_{\ce{Xe}}=\eta_{^{129}\text{Xe}}p_{\text{Xe}}/\left(k_{B}T_{\text{Xe}}\right)$,
where $\eta_{^{129}\text{Xe}}$ is the isotopic fraction of $^{129}\text{Xe}$
in the xenon gas. Every atom that undergoes two-photon excitation
will eventually emit an infrared photon, where a fraction $\epsilon_{T}$
of these photons will be detected, therefore the mean rate of infrared
photon detection is
\begin{equation}
\bar{n}_{f}=\frac{1}{2}W^{(2)}n_{\text{Xe}}\left(L\pi w_{0}^{2}\right)\epsilon_{T},\label{eq:average-fluorescence}
\end{equation}
where $\epsilon_{T}$ is the detection efficiency of the emitted infrared
photons, and the factor of $\frac{1}{2}$ accounts for the fact that
in an unpolarized atomic ensemble only half of the atoms get excited
(i.e., half of the ground-state atoms are in the $m=-1/2$ magnetic
sublevel while the other half are in the $m=+1/2$ sublevel). When
the atomic ensemble is polarized and subjected to an external magnetic
field, the infrared photon detection rate becomes
\begin{equation}
n_{f}(t)=\bar{n}_{f}\left[1+P_{\text{Xe}}\cos\left(\omega_{B}t\right)e^{-t/T_{d}}\right],\label{eq:fluorescence}
\end{equation}
where $T_{d}$ is the spin polarization's decay rate, $0\le P_{\text{Xe}}\le1$
is the initial degree of spin polarization.

\subsection{Detection Noise and Sensitivity Model}

\label{subsec:Detection-Noise-and}

The statistical uncertainty in the Larmor frequency detection arises
due to noise that accompanies the photodetector signal. Therefore,
the fractional sensitivity is $S=\sigma_{\omega_{B}}/\omega_{B}$,
where $\sigma_{\omega_{B}}^{2}$ is the total noise variance in the
measured Larmor frequency. The sensitivity is ultimately limited by
two fundamental sources of noise: shot noise and projection noise
\cite{budker_jackson-kimball_2013}. Shot noise results from the discrete
statistical nature of photons, described by Poisson statistics, with
variance $\sigma_{\omega_{B},SN}^{2}$. Projection noise is a result
of the Heisenberg uncertainty principle, and we label its variance
$\sigma_{\omega_{B},PN}^{2}$. The total noise variance is $\sigma_{\omega_{B}}^{2}=\sigma_{\omega_{B},SN}^{2}+\sigma_{\omega_{B},PN}^{2}$. The statistical uncertainty is reduced by averaging the results of
$N_{m}$ measurements of $\omega_{B}$ (step 4 in Section \ref{subsec:Experimental-Methodology}), thus the fractional sensitivity is given by
\begin{equation}
S=\sqrt{\frac{\sigma_{\omega_{B}}^{2}}{N_m\omega_{B}^{2}}}=\frac{1}{\sqrt{N_m}\omega_{B}}\sqrt{\sigma_{\omega_{B},SN}^{2}+\sigma_{\omega_{B},PN}^{2}}.\label{eq:sensitivity-single}
\end{equation}

The detected signal of Eq. \ref{eq:fluorescence} is a sum of a DC
component and a decaying sinusoidal, where we are interested in extracting
the sinusoidal's oscillation frequency, $\omega_{B}$. The variance
in the extracted frequency due to shot-noise is (see Appendix \ref{sec:Photon-shot-noise-v2})
\begin{equation}
\sigma_{\omega_{B},SN}^{2}=\frac{12T_{d}\left(\sqrt{1+\frac{P_{\text{Xe}}^{2}}{2}}\right)\left(e^{2T_{m}/T_{d}}-1\right)}{\bar{n}_{f}P_{\text{Xe}}^{2}T_{m}^{4}},\label{eq:shot_noise_variance}
\end{equation}
where $T_{m}$ is the time duration of a single measurement (step
3 in Section \ref{subsec:Experimental-Methodology}).

The variance due to spin projection noise in a single measurement
of the Larmor angular frequency is \cite{budker_jackson-kimball_2013}
\begin{equation}
\sigma_{\omega_{B},PN}^{2}=\frac{\left(2\pi\right)^{2}}{T_{d}N_{\text{Xe}}T_{m}},\label{eq:angular-frequency-projection-noise}
\end{equation}
where $N_{\text{Xe}}=\left(L\pi w_{0}^{2}\right)n_{\text{Xe}}$ is
the number of xenon atoms in the interaction volume.

Eqs. \ref{eq:sensitivity-single} - \ref{eq:angular-frequency-projection-noise} provide the ultimate fractional precision of the
proposed experiment for a given set of experimental parameter values.
This fractional sensitivity depends on several experimentally controlled
parameters, including the magnetic field strength and uniformity that
determine the Larmor frequency and spin polarization decay time; the
interaction volume, initial polarization, isotopic fraction of $\ce{^{129}Xe}$,
gas temperature, and gas pressure impact the decay time and the number
of interacting atoms; duration of one measurement and number of measurements
taken; and laser power, beam waist radius, and frequency. In the following
two sections we detail the values of these parameters.

\section{Natural and Constrained Parameter Values}

\label{sec:Parameter-Values}

In this section we provide details on parameter values that we use
in the calculation of the fractional sensitivity in the following
section. We detail atomic properties of $\ce{^{129}Xe}$, macroscopic
properties of xenon gas, characteristics of the magnetic field, laser
configuration, and explain how all these interact to impact the proposed
experimental measurement.

The electronic spectrum and physical properties of xenon isotopes
are well known. The reported gyromagnetic ratio of $\ce{^{129}Xe}$
is $\gamma=2\pi\times\SI{11.78e6}{Hz/T}$ \cite{Marshall-2020,Mohr-2012,Pfeffer-1994}.
The natural abundance of the isotope $\ce{^{129}Xe}$ is $\SI{26}{\percent}$
\cite{nikolaouNearunityNuclearPolarization2013a}. While isotopically
enriched xenon gas with significantly more than $\SI{80}{\percent}$$\ce{^{129}Xe}$
content is commercially available, here we conservatively assume $\eta_{^{129}\text{Xe}}=0.8$.
The $5\text{p}^{6}(^{1}\text{S}_{0})\rightarrow5\text{p}^{5}(^{2}\text{P}_{3/2})6\text{p}^{2}[5/2]_{2}$
two-photon transition is resonantly driven with photons at a DUV wavelength
of $\SI{256.01704(1)}{nm}$ \cite{Saloman-2004}. An excited xenon
atom will decay through one of two channels: $5\text{p}^{5}(^{2}\text{P}_{3/2})6\text{p}^{2}[5/2]_{2}\to5\text{p}^{5}(^{2}\text{P}_{3/2})6\text{s}^{2}[3/2]_{1}$
with a wavelength of $\SI{992.3}{nm}$ \cite{Saloman-2004} and a
splitting ratio of $0.637$ \cite{whiteheadDeactivationTwoPhoton1995},
or $5\text{p}^{5}(^{2}\text{P}_{3/2})6\text{p}^{2}[5/2]_{2}\to5\text{p}^{5}(^{2}\text{P}_{3/2})6\text{s}^{2}[3/2]_{2}$
with a wavelength of $\SI{904.5}{nm}$ \cite{Saloman-2004} and a
splitting ratio of $0.363$ \cite{whiteheadDeactivationTwoPhoton1995}
(see Fig. \ref{fig:experiment_and_energy_diagram}). By using circularly
polarized light, we can selectively excite only the $m=-1/2$ magnetic
sublevel of the ground state, so that population oscillation between
the two ground state magnetic sublevels is apparent in the detected
infrared fluorescence. Therefore, care must be taken to avoid also
driving the transition from the ground state $m=+1/2$ magnetic sublevel
to the $m=+5/2$ magnetic sublevel of the $F^{\prime}=5/2$ state
in the $5\text{p}^{5}(^{2}\text{P}_{3/2})6\text{p}^{2}[5/2]_{2}$
hyperfine manifold. The hyperfine splitting between the $F^{\prime}=5/2$
and $F^{\prime}=3/2$ state is $\Delta\omega_{hfs}=2\pi\times\SI{3.4105}{GHz}$
\cite{Xia-2010}. Below we show that the Doppler broadened absorption
profile FWHM is $\Delta\omega_{D}=2\pi\times\SI{1.26}{GHz}$, so $\Delta\omega_{D}<\Delta\omega_{hfs}$,
meaning we can selectively excite the transition to the $F^{\prime}=3/2$
state while avoiding excitation of the $F^{\prime}=5/2$ state with
a proper choice of laser parameters, as follows. We define the laser
cut-off frequency, $\omega_{l,\text{cut-off}}$, as the laser frequency
where the two-photon cross section of the $F^{\prime}=5/2$ state
is 10 time lower than the $F^{\prime}=3/2$ two-photon cross section,
i.e.,$\hat{\sigma}_{F=5/2}(\omega_{l,\text{cut-off}})/\hat{\sigma}_{F=3/2}(\omega_{l,\text{cut-off}})=0.1$.
Using Eq. \ref{eq:two-photon-cross-section}, this yields $\omega_{\text{cut-off}}=\omega_{\ce{Xe}}-1.18\Delta\omega_{D}$.
For all higher frequencies, this ratio is lower. Below we show that
$\omega_{\text{cut-off}}$ is outside the region of interest for practical
experiments, since the $F^{\prime}=3/2$ two-photon cross section
is too low so far off resonance. In fact, in the region of interest
$\hat{\sigma}_{F=5/2}/\hat{\sigma}_{F=3/2}\sim10^{-2}$. Therefore,
for all excitation frequencies of interest, the transition to the
$F^{\prime}=5/2$ state is neglected. Finally, there are multiple
reported values of the two-photon absorption coefficient, $\alpha_{\text{Xe}}$
\cite{raymondTwophotonLaserSpectroscopy1984,krollTwophotonAbsorptionPhotoionization1990a}.
Here we use the most recently reported value of $\SI{83(25)}{cm^{4}/J^{2}}$
\cite{krollTwophotonAbsorptionPhotoionization1990a}.

The Doppler FWHM is determined by natural properties of xenon (mass
and resonance frequency) and an experimentally controlled parameter,
i.e., gas temperature, with higher temperature resulting in a wider
Doppler width. For simplicity of the experimental apparatus and procedure,
we assume the gas is at room temperature, $T_{\ce{Xe}}=\SI{293.15}{K}$,
which results in a Doppler FWHM of $\Delta\omega_{D}=2\pi\times\SI{1.26}{GHz}$.
We note that precise control of this temperature is not required,
since precision of a few degrees K (commonly achieved with ordinary
room temperature control) would only change the excitation rate by
tiny fraction, and would not impact the measured Larmor frequency
and QG correction, which depend only on the applied magnetic field
strength and laser frequency.

Under the experimental conditions proposed here, Doppler broadening
dominates over all other broadening mechanisms. The natural linewidth
of the two-photon transition is about $\SI{4.3}{MHz}$ \cite{whiteheadDeactivationTwoPhoton1995}.
Collisional broadening has been measured to be $\SI{18.6}{MHz/mbar}$
\cite{raymondTwophotonLaserSpectroscopy1984}. Therefore, in order
to keep collisional broadening insignificant compared to Doppler broadening,
we need to keep the pressure much lower than $\SI{66}{mbar}$. Here
we consider a pressure of $\SI{10}{mbar}$, corresponding to collisional
broadening of $\SI{186}{MHz}$. Transit-time broadening width is given
by $\sim0.4v/w_{0}$ \cite{Demtroder2002}. At room temperature, the
most probable speed of a Boltzmann-Mawell distribution is $\sqrt{2k_{B}T/m_{\text{Xe}}}\approx\SI{200}{m/s}$,
where $m_{\text{Xe}}$ is the mass of $\ce{^{129}Xe}$. Therefore,
in order to keep Doppler broadening dominant over transit-time broadening,
we require $w_{0}\gg0.4\cdot2\pi v/\text{\ensuremath{\Delta\omega_{D}}}\approx\SI{63}{nm}$.
Even when considering speeds 10 times higher than the most probable
speed, we only need to keep the beam radius much greater than $\SI{0.63}{\mu m}$,
which is very easy to satisfy. As detailed below, we propose using
beam radii of at least $\SI{100}{\mu m}$, corresponding to a maximum
transit-time broadening of $\SI{8}{MHz}$ at 10 times the most probable
velocity. Compared to the Doppler width of $\SI{1.26}{GHz}$, all
of these broadening mechanisms make a negligible contribution of about
$\SI{200}{MHz}$ at most. Therefore, we neglect them and consider
only Doppler broadening.

The initial degree of ensemble spin polarization and its decay rate
have a strong impact on the sensitivity of the proposed measurement.
Methods for spin-polarizing $\ce{^{129}Xe}$ are well established
\cite{nikolaouNearunityNuclearPolarization2013a} and will not be
detailed here. We estimate the initial spin polarization, $P_{\text{Xe}}$,
based on published values in a systematic study of spin exchange optical
pumping of $\ce{^{129}Xe}$, where a polarization fraction of 90\%
has been experimentally achieved for gas pressure of $\SI{400}{mbar}$
\cite{nikolaouNearunityNuclearPolarization2013a}. While even higher
polarization is likely possible at the lower pressure considered here,
we conservatively assume $P_{\text{Xe}}=0.9$ for gas pressure of
$\SI{10}{mbar}$. Spin polarization can decay via depolarization or
through dephasing. Depolarization is also called longitudinal decay,
with a characteristic decay time $T_{1}$, and dephasing is commonly
termed transverse decay, with a characteristic timescale $T_{2}$.
Overall, spin polarization relaxation is dominated by $T_{2}$, since
any effect that reduces $T_{1}$ also reduces $T_{2}$. The spatial
uniformity of the magnetic field is critical in achieving long transverse
spin relaxation times. If different atoms experience a different magnetic
field, or equivalently the same atoms experience a varying magnetic
field as they perform diffusive motion in the gas container, then
their spin precession frequency will vary and they will become out
of phase with each other. Correspondingly, the transverse relaxation
time will decrease according to $T_{2}=2\pi/\gamma\Delta B$, where
$\Delta B$ is the spatial variation in the magnetic field across
the gas volume. However, the nature of the diffusive motion of the
atoms will change this dephasing rate, and may lead to a different
value of $T_{2}$, depending on gas pressure \cite{kilianFreePrecessionTransverse2007a}.
A recent experiment recorded $T_{1}$ and a $T_{2}$ values up to
$\SI{99}{hours}$ and $\SI{8000}{s}$, respectively, under a magnetic
field variation of $\SI{240}{pT}$ across a spherical container with
$\SI{2}{cm}$ diameter holding a 4:1 xenon:nitrogen gas mix with xenon
partial pressure of about $\SI{50}{mbar}$, where lower pressure correlated
with longer $T_{2}$ \cite{kilianFreePrecessionTransverse2007a}.
For these conditions, $2\pi/\gamma\Delta B\approx\SI{353}{s}$, which
is much smaller than the experimentally measured value. The difference
is attributed to the atoms' limited diffusive motion slowing down
the rate of dephasing. In another experiment, a field variance of
$\Delta B=\SI{800}{fT}$ over a distance of $\SI{2}{cm}$ in a nominal
magnetic field strength of $B=\SI{2.7}{\micro T}$ was demonstrated
\cite{Liu-2020}. We note that we obtained this value by using graph
extraction software to find the minimum field variation across $\SI{2}{cm}$
in the published graph of $\Delta B$ vs. spatial position. However,
this might be misleading, since this value is below the experimental
measurement precision, which is better than $\Delta B_{L}=\SI{3}{pT/cm}$.
Therefore, we conservatively assume a field variation of $\Delta B=\SI{6}{pT}$
over a sphere with $\SI{2}{cm}$ diameter. This corresponds to a transverse
relaxation time of $T_{2}=2\pi/\gamma\Delta B=\SI{3.93}{hours}$,
where we did not account for any diffusive effects that may increase
$T_{2}$ further. The Larmor frequency corresponding to a $\SI{2.7}{\micro T}$
magnetic field is $2\pi\times\SI{31.8}{Hz}$. We note that the coil
system used to produce this field provides plenty of room and clearance
(more than $\SI{50}{cm}$ on each side) for the gas container, with
optical access on six sides that can accommodate the excitation laser
beam and fluorescence collection optics for the detection system in
the experiment proposed here.

The power of the two-photon excitation laser has a strong impact on
the fractional sensitivity. As noted above, the excitation wavelength
of $\SI{256}{nm}$ is in the DUV, where achieving high laser power
and frequency stability are technically challenging. The established
approach for addressing this challenge is through frequency upconversion
via cascaded nonlinear wave mixing of near-infrared lasers, either
CW or ultrafast frequency combs. Since the efficiency of nonlinear
wave mixing strongly scales with intensity, single-pass upconversion
of CW lasers suffers from low efficiency, producing very limited DUV
power. Instead, researchers perform the upconversion process inside
a cavity, where very high optical power is built up and the laser
light effectively passes through the nonlinear medium many times \cite{Sayama-1998}.
For example, Burkley et al. have generated $\SI{1.4}{W}$ at $\SI{243.1}{nm}$
with a linewidth $<\SI{10}{kHz}$ using a $\SI{10}{W}$ infrared seed
laser \cite{burkleyHighlyCoherentWattlevel2019}. Another upconversion
method uses frequency combs, which are frequency-stabilized pulsed
lasers. Frequency combs usually operate with MHz-GHz pulse repetition
rate, where pulse-to-pulse coherence leads to a set of very narrow
spectral features (``comb teeth''), equally spaced at intervals
equal to the pulse repetition rate \cite{fortier20YearsDevelopments2019}.
Their short pulse duration (on the femtosecond-picosecond scale) leads
to very high intensity and efficient single-pass frequency conversion.
For example, Yang et al. \cite{YangUVcomb} demonstrated a DUV frequency
comb at $\SI{258}{nm}$ that boasts a sub-Hz linewidth and delivers
$\SI{1.58}{W}$ of power, produced from a $\SI{41}{W}$ infrared frequency
comb seed laser. More recently, phase-stabilized femtosecond lasers
with wavelength near $\SI{1}{\micro m}$ have reached $\SI{1}{kW}$
power \cite{Shestaev_kW_phase_stable}, and a similar system yielded
$\SI{10.4}{kW}$ power \cite{Muller-2020}. At the same time, frequency
conversion from such high-power sources has also been advancing: Rothhardt
et al. have demonstrated a femtosecond laser delivering $\SI{100}{W}$
at $\SI{343}{nm}$ from a $\SI{620}{W}$ infrared seed laser \cite{rothhardt100AveragePower2016}.
Therefore, here we consider a DUV laser power range of $\SI{1}{W}\text{ to }\SI{100}{W}$,
where the laser can be tuned across the Doppler lineshape, and the
laser linewidth is $\ll\Delta\omega_{D}$. These laser properties
have either been demonstrated or have strong evidence for their feasibility
with existing or near-future technology.

An additional constraint must be considered when a frequency comb
laser is considered for the proposed experiment. When a frequency
comb excites a two-photon transition, comb teeth that are equally
spaced around the resonance frequency combine coherently to drive
the transition, resulting in a transition rate equivalent to a CW
laser of the same average power \cite{baklanovTwophotonAbsorptionUltrashort1977,PiqueNetPhot}.
However, here we are proposing to use a comb to selectively excite
a subset of atoms, with a particular velocity, out of a large number
of atoms with a wide velocity distribution, while also avoiding the
transition to the $F^{\prime}=5/2$ state. In other words, we wish
to excite only a narrow spectral span out of the Doppler broadened
spectrum of the transition to the $F^{\prime}=3/2$ state. If the
comb's pulse repetition rate (equal to its spectral ``teeth'' spacing)
is smaller than the Doppler width, then sets of comb teeth will combine
to excite different velocity subsets of atoms. Also, if the pulse
repetition rate is close to the hyperfine splitting (i.e., no more
than a Doppler width away), then the transitions to both the $F^{\prime}=3/2$
and $F^{\prime}=5/2$ states get excited. Therefore, in order to be
able to excite a single velocity group to the $F^{\prime}=3/2$ state
alone, the comb's repetition rate must significantly exceed the combined
hyperfine splitting and Doppler width. We consider a pulse repetition
rate, $f_{\text{rep}}$, that is triple the Doppler FWHM plus the
hyperfine splitting to be sufficient, i.e., $2\pi f_{\text{rep}}\ge3\Delta\omega_{D}+\Delta\omega_{hfs}$.
Under this condition, even in the extreme case where the comb is tuned
off the $F^{\prime}=3/2$ resonance by a full FWHM, to $\omega_{\text{Xe}}+\Delta\omega_{D}$,
the next nearest resonance is in the $F^{\prime}=5/2$ level, where
the excitation's cross section is $2.5\times10^{-4}$ times smaller,
and therefore can be neglected. For the case considered here, our
condition translates into a pulse repetition rate of at least $\SI{7.2}{GHz}$,
which is commonly achieved in various near-infrared comb systems \cite{bartels10GHzSelfReferencedOptical2009,Kippenberg-2011}.
We note that increasing the pulse repetition rate reduces pulse energy
for the same average power, however it makes ultrafast pulse amplification
to high power more straight-forward since pulse-distorting parasitic
nonlinear effects are better mitigated \cite{Agrawl-2012,ZervasHighPowerFiberLaserReview,LimpertUltrafastFiberLaserSystems}.
Overall, the lower pulse energy and shorter pulse duration roughly
balance out, delivering similar peak intensity, and therefore similar
nonlinear upconversion efficiency downstream. Additionally, the repetition
rate of a comb can be increased from the MHz to the GHz range using
an external Fabry-Perot cavity with high transmission \cite{Steinmetz-2009,Mildner-2016,Lesundak-2015,HouGHzComb}.
Therefore, the required pulse repetition rate is well within reach
of existing technology.

Finally, the fractional sensitivity depends on the detection efficiency
of the emitted infrared photons, $\epsilon_{T}=\epsilon_{det}\epsilon_{geo}$,
where $\epsilon_{det}$ is the efficiency of the detector and $\epsilon_{geo}$
is the fraction of fluorescence photons that are collected into the
detector. As noted above, the emission is in the wavelengths of $\SI{992}{nm}$
and $\SI{905}{nm}$. These are in the near-infrared and therefore
can be detected with a high efficiency, e.g., using an avalanche photodetector.
$\epsilon_{geo}$ depends on the collections optics geometry, whose
details are out of the scope of the present work. Standard methods
commonly achieve $\sim50\%$ collection efficiency and $\sim45\%$
detection efficiency, for an overall efficiency of $\sim22.5\%$ \cite{Cerez_collection_efficiency}.
Here we conservatively assume an overall efficiency of $\epsilon_{T}=0.1$,
which incorporates the efficiency of both the detector and the photon
collection geometry.

A summary of all parameter values are shown in Table \ref{tab:lit-values-1}.
The values that are calculated from these parameters are shown in
Table \ref{tab:calc-values}, some of which were obtained via the
optimization procedure detailed in Section \ref{sec:results}.

\begin{table*}
\begin{ruledtabular}
\caption{Parameter values estimated from published literature.}
\label{tab:lit-values-1}
\begin{tabular}{llll}
Name & Symbol & Value & Ref\tabularnewline
\hline
Laser Power limits & $P_{l}$ & $\SI{1}{W}\text{ to }\SI{100}{W}$ & See Section \ref{sec:Parameter-Values}\tabularnewline
Two-photon transition resonant wavelength & $\lambda_{\text{Xe}}$ & $\SI{256.01704(1)}{nm}$ & \cite{Saloman-2004}\tabularnewline
Two-photon transition resonant angular frequency & $\omega_{\text{Xe}}$ & $\SI{7.3575242(3)e15}{1/s}$ & \cite{Saloman-2004}\tabularnewline
$\ce{^{129}Xe}$ nuclear spin gyromagnetic ratio & $\gamma$ & $2\pi\times\SI{11.78e6}{Hz/T}$ & \cite{Marshall-2020,Mohr-2012,Pfeffer-1994}\tabularnewline
Two-photon transition absorption coefficient & $\alpha_{\text{Xe}}$ & $\SI{83}{cm^{4}/J^{2}}$ & \cite{raymondTwophotonLaserSpectroscopy1984}\tabularnewline
Magnetic field & $B$ & $\SI{2.7}{\micro T}$ & \cite{Liu-2020}\tabularnewline
Variation in the magnetic field & $\Delta B$ & $\SI{6}{pT}$ & \cite{Liu-2020}\tabularnewline
Xenon gas temperature & $T_{\text{Xe}}$ & $\SI{300}{K}$ & \cite{kilianFreePrecessionTransverse2007a}\tabularnewline 
Xenon gas pressure & $p_{\text{Xe}}$ & $\SI{10}{mbar}$ & \cite{kilianFreePrecessionTransverse2007a,nikolaouNearunityNuclearPolarization2013a}\tabularnewline
Isotopic fraction of $\ce{^{129}Xe}$ & $\eta_{^{129}\text{Xe}}$ & $\SI{80}{\%}$ & \cite{nikolaouNearunityNuclearPolarization2013a}\tabularnewline 
Initial spin polarization & $P_{\text{Xe}}$ & $\SI{90}{\%}$ & \cite{nikolaouNearunityNuclearPolarization2013a}\tabularnewline 
Interaction length & $L$ & $\SI{2}{cm}$ & \cite{nikolaouNearunityNuclearPolarization2013a,kilianFreePrecessionTransverse2007a}\tabularnewline 
Total efficiency of fluorescence detection & $\epsilon_{T}$ & $\SI{10}{\%}$ & \cite{Cerez_collection_efficiency}\tabularnewline 
\end{tabular}
\end{ruledtabular}
\end{table*}

\begin{table*}
\begin{ruledtabular}
\caption{Values calculated from parameters from Table \ref{tab:lit-values-1}.}
\label{tab:calc-values}
\begin{tabular}{lll}
Name & Symbol & Value\tabularnewline
\hline 
Larmor frequency without QG & $\omega_{B}$ & $2\pi\times\SI{31.78}{Hz}$\tabularnewline
Laser beam radius & $w_{0}$ & $\SI{100}{\micro m}$ to $\SI{999}{\micro m}$\tabularnewline 
Spin relaxation time & $T_{d}$ & $\SI{3.93}{hours}$\tabularnewline
Doppler profile FWHM & $\Delta\omega_{D}$ & $2\pi\times\SI{1.27}{GHz}$\tabularnewline
Duration of a single measurement & $T_{m}$ & $\SI{4.67}{hours}$\tabularnewline
Optimal laser frequency & $\omega_{l,opt}$ & $\SI{7.3575378e15}{1/s}$\tabularnewline
\end{tabular}
\end{ruledtabular}
\end{table*}

\section{Sensitivity Optimization, Detection Bounds and Discussion}

\label{sec:results}

\subsection{Optimizable Parameter Values}

With the experimental parameter values detailed in the previous section
there are still five values that have not been given a specified value:
$w_{0}$, $T_{m}$, $N_{m}$, $\omega_{l}$, and $P_{l}$. We treat
these in this section.

The interaction between the xenon atoms, the magnetic field, and the
laser light occurs in a volume defined by the interaction length,
$L$, and the beam waist diameter, $w_{0}$ \textbf{.} We have assumed
that the interaction length is much shorter than the beam's confocal
parameter and therefore we take the beam diameter to be constant along
$L$. In the calculations carried out below, the beam waist radius
is optimized for each value of laser power by setting it to the value
that minimizes the fractional sensitivity of Eq. \ref{eq:sensitivity-single},
as derived in Appendix \ref{sec:Optimizing_w0}. When the laser is
on resonance with the two-photon transition and with laser power values
ranging from $\SI{1}{W}$ to $\SI{100}{W}$, the optimal beam radii
range from $\SI{100}{\micro m}$ to $\SI{999}{\micro m}$, with corresponding
confocal parameter of $\SI{24}{cm}$ to $\SI{24}{m}$, which satisfies
our requirement that $b\gg L=\SI{2}{cm}$.

The power of the laser, $P_{l}$, and the number of measurement iterations,
$N_{m}$, will unambiguously give a more precise result as they increase. As explained
in the preceding section, we constrain the laser power to feasible
values of $\SI{1}{}$-$\SI{100}{W}$. Below, we keep $N_{m}$ bounded
by a total experimental measurement time of 1 year, in order to keep
the proposed experiment within the bounds of feasibility. The remaining
two variables are the duration of a single experiment, $T_{m}$, and
the laser angular frequency, $\omega_{l}$. Next, we optimize them
to numerically yield minimal values of the fractional sensitivity
of Eq. \ref{eq:sensitivity-single} .

The goal of optimizing the value of $T_{m}$ is to find the duration
of a single measurement that delivers a given sensitivity with the
minimal duration of an entire set of measurements, $N_{m}T_{m}$.
To find this optimal value we calculated $S$ with different $P_{l}$
and $\omega_{l}$ while sweeping over a range of values for $T_{m}$
and finding the number of experiments required to achieve a specific
value of $S$. For all considered values of $P_{l}$ and $\omega_{l}$,
which were $\SI{1}{}$-$\SI{100}{W}$ and $\omega_{Xe}-1.75\Delta\omega_{D}$
to $\omega_{Xe}+1.75\Delta\omega_{D}$, respectively, the optimal
value which minimizes $T_{m}N_{m}$ was found to be $T_{m}=1.19T_{d}$
in all cases. This value of $T_{m}$ was used in all the calculations
below.

\begin{figure}
\includegraphics[width=0.45\textwidth]{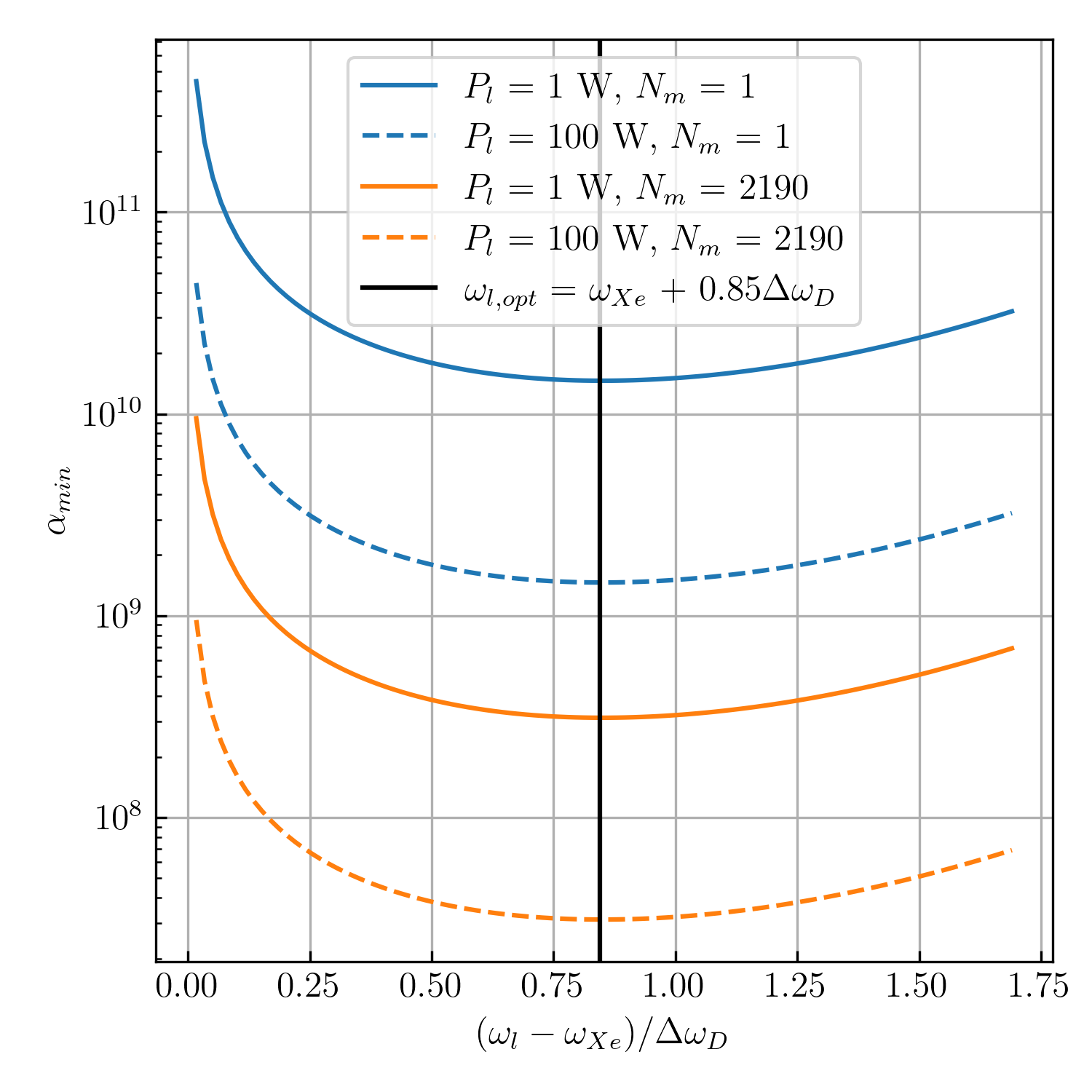}
\caption{Minimum detectable QG correction coefficient, $\alpha_{0,min}$, as
a function of laser frequency, $\omega_{l}$, for different values
of laser power, $\omega_{l}$, and number of measurements, $N_{m}$.
Four examples are shown with a combination of two values for ($\SI{1}{W}$
and $\SI{100}{W}$) and two values for $N_{m}$ (1 and 2190). The
latter correspond total measurement times, $N_{m}T_{m}$, of $\SI{4.7}{hours}$
and $\SI{1}{year}$, respectively. The optimal laser frequency, $\omega_{l,opt}=\omega_{\ce{Xe}}+0.85\Delta\omega_{D}$,
is identified by a black vertical line.}
\label{fig:opt_w_l}
\end{figure}

In optimizing $\omega_{l}$ we considered the case where one measurement
is taken on two-photon resonance and another measurement is taken
at a laser frequency that minimizes the value of $\alpha_{0}$ that
can be detected for given values of $P_{l}$ and $N_{m}$, i.e., we
find the minimum value of $\alpha_{0,min}=S\left(\omega_{l},P_{l},N_{m}\right)\left(M_{p}c/mv\right)$
for this set of $P_{l}$ and $N_{m}$ values. There are two opposing
factors influencing the value of $\alpha_{0,min}$ as $\omega_{l}$
is varied. The first factor is the decreased velocity of the atoms
that are being probed as $\omega_{l}$ approaches $\omega_{\text{Xe}}$,
meaning a smaller $v$ and therefore higher $\alpha_{0,min}$. The
second factor is the increased Doppler-broadened two-photon cross-section
as $\omega_{l}$ approaches $\omega_{\text{Xe}}$, resulting in a
higher two-photon transition rate and therefore smaller $S$ and lower
$\alpha_{0,min}$. These opposing trends produce a minimum in $\alpha_{0,min}$,
which we find numerically, where $S$ is given by Eq. \ref{eq:sensitivity-single}
and $v$ is given by Eq. \ref{eq:lambda_and_velocity}. For all values
of $P_{l}$ and $N_{m}$ that we checked, the minimum was always at
$\omega_{l,opt}=\omega_{\text{Xe}}+0.85\Delta\omega_{D}$. A few examples
are shown in Fig. \ref{fig:opt_w_l}. Note that the curvature of the
curves in Fig. \ref{fig:opt_w_l} is small, meaning that deviating
from the optimal frequency does not greatly increase the value of
$\alpha_{0,min}$. For example, using $\omega_{l}=\omega_{\text{Xe}}+0.5\Delta\omega_{D}$
would increase $\alpha_{0,min}$ by no more than 30\%. This implies
that adding more measurement points at different laser frequencies
would not change the order of magnitude of the duration of the experiment,
and therefore is still feasible.

\subsection{Calculation Results}

\begin{figure}
\includegraphics[width=0.45\textwidth]{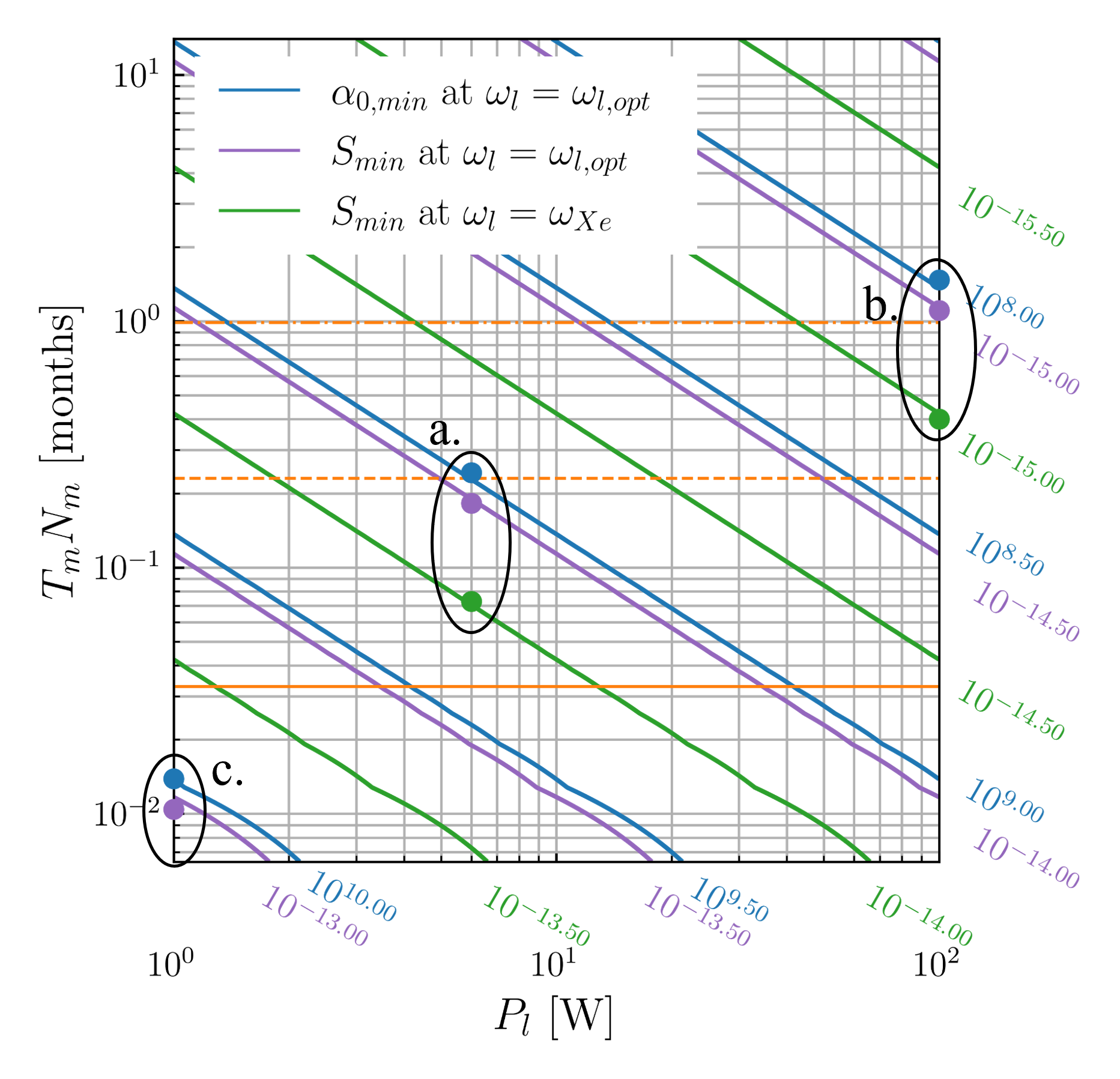}
\caption{The minimum values of the QG correction coefficient $\alpha_{0,min}$
at the optimal laser frequency $\omega_{l,opt}$ (blue), the corresponding
optimized fractional sensitivity $S_{min}$ at $\omega_{l}=\omega_{,opt}$
(purple), and the optimized fractional sensitivity on two-photon resonance,
$S_{min}$ at $\omega_{l}=\omega_{\ce{Xe}}$ (green). Notable time
scales of 1 day (solid), week (dashed), and month (dot-dashed) are
indicated in orange. }
\label{fig:alpha_omega_l_opt}
\end{figure}

The calculation results are shown in Fig. \ref{fig:alpha_omega_l_opt}
as three superimposed contours as a function of laser power and the
total measurement time for a single laser frequency, $T_{m}N_{m,}$.
For convenience the plot includes horizontal lines that mark total
times of one day, one week, and one month. The three overlayed contours
depict $\alpha_{0,min}$ for $\omega_{l}=\omega_{l,opt}$, the corresponding
optimized fractional sensitivity, $S_{min}$, and the value of $S_{min}$
for $\omega_{l}=\omega_{\ce{Xe}}$.

Fig. \ref{fig:alpha_omega_l_opt} is an aid to determine the feasibility
of detecting a given value of the QG correction coefficient, $\alpha_{0}$,
for a given laser power, by establishing the duration of the required
measurement. This is done in three steps. First, determine the total
measurement time, $T_{m}N_{m}$, to detect the specified value of
$\alpha_{0}$ with the available laser power and with the laser frequency
tuned to its optimal value, i.e., $\omega_{l}=\omega_{l,opt}$. Next,
determine the corresponding optimized fractional sensitivity, $S_{min}$
at $\omega_{l}=\omega_{l,opt}$. Finally, determine the total measurement
time to achieve the same fractional sensitivity when the laser is
tuned to the two-photon resonance, i.e., $\omega_{l}=\omega_{Xe}$.

We next provide a few examples that illustrate the feasibility of
measuring various values of $\alpha_{0}$. The first example is indicated
in Fig. \ref{fig:alpha_omega_l_opt} by the three markers in the ellipse
marked ``a'', which lie along the vertical line corresponding to
a laser power of $P_{l}=\SI{6}{W}$. The blue marker indicate the
a target value of $\alpha_{0}\approx10^{9}$, where the required time
to measure the Larmor frequency with adequate precision using a laser
frequency of $\omega_{l}=\omega_{l,opt}$ is approximately 1 week.
This corresponds to a fractional sensitivity of $S\approx10^{-14}$
(purple marker). To achieve the same fractional sensitivity at resonance,
i.e., with a laser frequency $\omega_{l}=\omega_{\ce{Xe}}$, a total
time of $T_{m}N_{m}\approx\SI{2}{days}$ is required (green marker).
This means that the whole experiment will take about 1.5 weeks, which
is feasible. Taking additional data points at other laser frequencies
is also feasible, since each would take $\sim9$ days of measurement
(see discussion of Fig. \ref{fig:opt_w_l}). This example shows that
detecting $\alpha_{0}\approx10^{9}$ is feasible with just a minor
improvement to existing laser technology.

As another example, indicated by the markers in the ellipse marked
``b'' in Fig. \ref{fig:alpha_omega_l_opt}, consider an experiment
using a laser with $P_{l}=\SI{100}{W}$ and a target of $\alpha_{0}\approx10^{8}$.
With a laser frequency $\omega_{l}=\omega_{l,opt}$ this requires
a measurement time of $\sim6$ weeks, where the corresponding sensitivity
is $S\approx10^{-15}$. To achieve this sensitivity on resonance would
take $\sim2$ weeks, for a total measurement time of 2 months, where
each additional laser frequency data point needs $\sim8$ weeks to
complete. While this is still feasible, such an experiment would be
a very challenging task, even for the extrapolated DUV laser power
of $P_{l}=\SI{100}{W}$ that we expect to be available in the near
future.

For a final example, consider an experiment using a laser with $P_{l}=\SI{1}{W}$
and a target of $\alpha\approx10^{10}$, corresponding to the markers
in the ellipse marked ``c'' in Fig. \ref{fig:alpha_omega_l_opt}.
The measurement with a laser frequency $\omega_{l}=\omega_{l,opt}$
will take about 12 hours and have a sensitivity of $S\approx10^{-13}$.
For on-resonance measurement, this would take about 3 hours (less than a single optimized measurement time $T_m$). This gives a total
measurement time of about $\SI{15}{hours}$ for two data points, with
additional points taking only $\sim16$ hours each. This example shows
that detecting $\alpha_{0}\approx10^{10}$ is within reach of existing
technology.

\section{Conclusion}

In conclusion, we proposed using two-photon spectroscopy of a thermal
ensemble of hyperpolarized $\ce{^{129}Xe}$ atoms to measure theoretically
predicted QG-induced Larmor frequency variation. We quantitatively
demonstrated the feasibility of this approach, where existing or near-future
technology would provide sufficient precision to go beyond known bounds.
Key technical aspects are a magnetic field with a high degree of spatial
uniformity that leads to long coherence time of $\ce{^{129}Xe}$ nuclear
spin precession, and a high power DUV laser that can drive a two-photon
transition at a high rate. Particularly, we modeled the detected signal
and detection noise, culminating in an analytical expression that
describes the fractional sensitivity in terms of the experimental
parameters. We optimized the experimental parameters to detect the
QG correction that is linear in momentum, parametrized by a coefficient
$\alpha_{0}$, which existing experimental evidence bound be less
than $10^{17}$ . Finally, we discuss the feasibility of detecting
various values of $\alpha_{0}$, showing that $10^{10}$ is within
reach of current technology, while $10^{8}$ would be feasible with
near-future laser capabilities, corresponding to improving the current
bound by 7 and 9 orders of magnitude, respectively.

\begin{acknowledgments}
We would like to thank Dr. Lindsay LeBlanc for helpful discussions
and her support as a co-supervisor of James Maldaner. We acknowledge
the support of the Natural Sciences and Engineering Research Council
of Canada (NSERC), {[}funding reference number RGPIN-2019-05017{]}.
We acknowledge the support of the Government of Canada\textquoteright s
New Frontiers in Research Fund (NFRF), {[}NFRFE-2018-01220{]}. We
acknowledge the support of Alberta Innovates {[}570922-21 and 212200789{]}. We acknowledge the support of Alberta's Quantum Technologies Major Innovation Fund {[}RCP-19-004-MIF. We acknowledge the support of Quantum City {[}1059845-10{]}. 
\end{acknowledgments}

\appendix

\section{Two-photon cross-section}

\label{sec:Two-photon-cross-section}

Here we develop an analytical expression for the two-photon cross
section, and make connections to parameters whose values are reported
in literature. The two-photon cross-section is the product of three
values: a photon statistics factor of the light source, $G^{\left(2\right)}$,
the lineshape of the two-photon transition, $g\left(\omega\right)$,
and the two-photon cross-section's value at line center, $\sigma_{0}^{\left(2\right)}$,

\begin{equation}
\hat{\sigma}^{(2)}\left(\omega\right)=\sigma_{0}^{(2)}g\left(\omega\right)G^{(2)},
\end{equation}
where $\omega$ is the angular frequency of the radiation.

We take $G^{(2)}=1$ for a classically coherent light source. Here,
Doppler broadening dominates the lineshape so we take it to be a Gaussian
that is defined by the Doppler FWHM $\Delta\omega_{D}$. Published
reports of two-photon cross-section measurements commonly write the
Gaussian lineshape using a width parameter, $\beta$,

\begin{equation}
g(\omega)=\exp\left[-\beta\left(\omega-\omega_{0}\right)^{2}\right],
\end{equation}
where $\beta=\ln2\left(\frac{2}{\Delta\omega_{D}}\right)^{2}$ and
the cross section is maximized at $\omega_{0}$. Additionally, the
reported value is usually the the two-photon absorption coefficient,
$\alpha=\sigma_{tot}/(\hbar\omega)^{2}$, where $\sigma_{\text{tot}}$
is spectrally integrated two-photon cross-section given by
\begin{equation}
\sigma_{tot}=\int_{-\infty}^{\infty}\sigma_{0}^{(2)}g\left(\omega\right)\,d\omega=\sigma_{0}^{(2)}\sqrt{\frac{\pi}{\beta}}.
\end{equation}

We found it most convenient to express the two-photon cross-section
using $\alpha$ and $\Delta\omega_{D}$,
\begin{equation}
\begin{split}
\hat{\sigma}^{(2)}(\omega)=&\alpha\left(\hbar\omega\right)^{2}\sqrt{\frac{\ln2}{\pi}}\frac{2}{\Delta\omega_{D}}\cdot\\&\exp\left[-\ln2\left(\frac{2}{\Delta\omega_{D}}\right)^{2}\left(\omega-\omega_{0}\right)^{2}\right]G^{(2)}.
\end{split}
\end{equation}
In the main text the incident radiation is denoted $\omega=\omega_{l}$
and the resonant frequency is $\omega_{0}=\omega_{\ce{Xe}}$.

\section{Photon shot noise}

\label{sec:Photon-shot-noise-v2}

The variance in the estimation of the frequency of an exponentially
decaying sinusoidal signal, due to shot noise in the signal's amplitude,
is \cite{swallowsPermDipoleMoment}
\begin{equation}
\sigma_{\omega}^{2}=\frac{6n^{2}\left(e^{2\Gamma T_{m}}-1\right)}{\Gamma A^{2}T_{m}^{4}},\label{eq:Swallows_variance}
\end{equation}
where $n$ is the RMS noise spectral density amplitude at the frequency
of the sinusoid, $\Gamma$ is the decay rate of the signal, $A$ is
the initial amplitude of the signal, and $T_{m}$ is the duration
of the time interval taken into account for estimating the frequency.

Here, the signal is the oscillating part of the photodetector photocurrent
corresponding to the infrared photon detection rate of Eq. \ref{eq:fluorescence}.
The detected photocurrent is
\begin{equation}
\begin{split}
i & =en_{f}=e\bar{n}_{f}\left[1+P_{Xe}\cos\left(\omega_{B}t\right)e^{-t/T_{d}}\right]\\&=i_{DC}+i_{AC}e^{-t/T_{d}},
\end{split}
\end{equation}
where $e$ is the electron charge, and $i_{DC}=e\bar{n}_{f}$ and
$i_{AC}=e\bar{n}_{f}P_{Xe}\cos\left(\omega_{B}t\right)$ are its DC
and non-decaying AC components, respectively. Therefore, $T_{m}$
is the time duration of a single measurement, $\Gamma=1/T_{d}$, and
$A=\max\left(i_{AC}\right)=e\bar{n}_{f}P_{Xe}$. The RMS noise spectral
density amplitude in $i_{n}=i_{DC}+i_{AC}$ due to shot noise collected
in a bandwidth of $B_{n}$ is $n=\sqrt{\sigma_{i_{n}}^{2}/B_{n}}$,
where $\sigma_{i_{n}}^{2}=2ei_{n,RMS}B_{n}$ is the shot noise variance
in a photocurrent with RMS value of $i_{n,RMS}$, integrated over
a bandwidth $B_{n}$. Overall, $n=\sqrt{2ei_{n,RMS}}$, where
\begin{equation}
i_{n,RMS}=\sqrt{i_{DC,RMS}^{2}+i_{AC,RMS}^{2}}=e\bar{n}_{f}\sqrt{1+\frac{P_{Xe}^{2}}{2}}.
\end{equation}
 Substituting the expressions for $\Gamma$, $A$ and $n$ into Eq.
\ref{eq:Swallows_variance}, we obtain
\begin{equation}
\sigma_{\omega_{B},SN}^{2}=\frac{12T_{d}\left(\sqrt{1+\frac{P_{Xe}^{2}}{2}}\right)\left(e^{2T_{m}/T_{2}}-1\right)}{\bar{n}_{f}P_{Xe}^{2}T_{m}^{4}}
\end{equation}
for the shot noise variance in the estimation of the frequency of
the detected photocurrent signal.

\section{Optimization of the laser beam radius }

\label{sec:Optimizing_w0}

Using the analytical model of Eq. \ref{eq:sensitivity-single}, we
want to determine the laser beam radius, $w_{0}$, that minimizes
the fractional sensitivity. We begin by finding an expression for
$w_{0}$ that corresponds to $\partial S/\partial w_{0}=0$, as follows.
First, we write $S^{2}$ in a manner that shows its dependence on
$w_{0}$,
\begin{align}
S^{2} & =\frac{1}{\omega_{B}^{2}N_{m}}\left(A_{SN}w_{0}^{2}+\frac{A_{PN}}{w_{0}^{2}}\right),
\end{align}
where we defined
\begin{equation}
\begin{split}
A_{SN}  =&\frac{12T_{2}\left(1+\frac{P_{Xe}^{2}}{2}\right)^{1/2}}{\alpha\sqrt{\frac{\ln2}{\pi}}\frac{2}{\Delta\omega_{D}}\left(\frac{2P_{l}}{\pi}\right)^{2}\frac{1}{2}n_{Xe}L\epsilon_{f}\pi P_{Xe}^{2}T_{m}^{4}}\cdot\\
&\frac{(e^{2T_{m}/T_{2}}-1)}{\exp\left[-4\ln2\left(\frac{\omega_{l}-\omega_{0}}{\Delta\omega_{D}}\right)^{2}\right]}
\end{split}
\end{equation}
and
\begin{equation}
A_{PN}=\frac{2\pi}{T_{2}n_{Xe}L\pi T_{m}}.
\end{equation}
Next, we differentiate the equation
\begin{align}
\omega_{B}^{2}N_{m}S^{2} & =A_{SN}w_{0}^{2}+A_{PN}\frac{1}{w_{0}^{2}}
\end{align}
and obtain
\begin{align}
\omega_{B}^{2}N_{m}\left(2\frac{\partial S}{\partial w_{0}}S\right) & =A_{SN}2w_{0}-A_{PN}2\frac{1}{w_{0}^{3}}.\label{eq:diff_for_opt_w0}
\end{align}
Finally, we set $\frac{\partial S}{\partial w_{0}}=0$ and solve for
$w_{0}$, which yields
\begin{equation}
w_{0}=\left(\frac{A_{PN}}{A_{SN}}\right)^{1/4}.
\end{equation}
Differentiating Eq. \ref{eq:diff_for_opt_w0} produces
\begin{align}
2\omega_{B}^{2}N_{m}\left[\frac{\partial^{2}S}{\partial w_{0}^{2}}S+\left(\frac{\partial S}{\partial w_{0}}\right)^{2}\right] & =2A_{SN}+6A_{PN}\frac{1}{w_{0}^{4}},
\end{align}
which we evaluate for $w_{0}=\left(\frac{A_{PN}}{A_{SN}}\right)^{1/4}$
where $\frac{\partial S}{\partial w_{0}}=0$, so
\begin{align}
2\omega_{B}^{2}N_{m}\left[\frac{\partial^{2}S}{\partial w_{0}^{2}}S\right]_{w_{0}=\left(\frac{A_{PN}}{A_{SN}}\right)^{1/4}} & =8A_{SN}.
\end{align}
All variables in this equation are positive, therefore $\left[\frac{\partial^{2}S}{\partial w_{0}^{2}}\right]_{w_{0}=\left(\frac{A_{PN}}{A_{SN}}\right)^{1/4}}$
is positive, and the extremum at $w_{0}=\left(\frac{A_{PN}}{A_{SN}}\right)^{1/4}$
is a minimum. This expression for $w_{0}$ is used in all calculations
in this paper.

\end{document}